\documentclass[oneside,reqno]{article}
\usepackage{tabularx,longtable,booktabs,multirow,graphicx,lscape,amsmath}
\usepackage[numbers, super, sort&compress]{natbib}
\usepackage{hyperref}
\usepackage[nolist]{acronym}
\usepackage[normalem]{ulem}
\usepackage[usenames, dvipsnames]{color}
\usepackage[table, xcdraw]{xcolor}
\usepackage[margin=1.5in]{geometry}
\usepackage{soul}
\usepackage{rotating}
\usepackage{arydshln}
\usepackage{rotating}
\usepackage{setspace}

\definecolor{mycolor1}{RGB}{220,220,255}
\definecolor{mycolor2}{RGB}{0, 0, 225}

\newcommand{\insertitem}[1]{%
    \begin{center}
        \vspace{10pt} 
        [INSERT #1 HERE.]
        \vspace{10pt} 
    \end{center}
}
\newcommand{\hypothesis}[1]{
    \begin{center}
        \textit{#1}
    \end{center}
}

\begin{document}
\begin{acronym}[MPC]
    \acro{2SLS}{2-Stage Least Squares}
    \acro{AMI}{Acute Myocardial Infarction}
    \acro{CDHP}{Consumer-Driven Health Plans}
    \acro{CMS}{Centers for Medicare and Medicaid Services}
    \acro{VM}{Value-Based Payment Modifiers Program}
    \acro{COB}{Coordination of Benefits}
    \acro{DiD}{Difference-in-Difference}
    \acro{DRGs}{Diagnosis Related Groups}
    \acro{EP}{Eligible Professional}
    \acro{FDA}{Food and Drug Administration}
    \acro{FTC}{Federal Trade Commission}
    \acro{GNT}{Gowrisankaran, Nevo, and Town}
    \acro{GP}{General Practitioners}
    \acro{HHI}{Herfindahl-Hirschman Index}
    \acro{HMO}{Health Maintenance Organizations}
    \acro{HMOs}{Health Maintenance Organizations}
    \acro{HRR}{Hospital Referral Regions}
    \acro{HSA}{Health Service Areas}
    \acro{MarketScan data}{Truven Health MarketScan\textsuperscript{\textregistered} Research Databases}
    \acro{MSA}{Metropolitan Statistical Area}
    \acro{NHS}{National Health Service}
    \acro{OOP}{out-of-pocket}
    \acro{pdf}{probability density function}
    \acro{PPO}{Preferred-Provider Organizations}
    \acro{ProMedica}{ProMedica Health System, Inc.}
    \acro{SCM}{Synthetic Control Method}
    \acro{SCP}{Structure-Conduct-Performance}
    \acro{SD}{Standard Deviation}
    \acro{SLH}{St. Luke's Hospital}
    \acro{SOI}{Severity of Illness}
    \acro{Truven Health}{Truven Health Analytics\textsuperscript{\textregistered}, part of the IBM Watson Health\texttrademark business}
    \acro{UK}{United Kingdom}
    \acro{WTP}{willingness to pay}
    \acro{FE}{fixed effects}
    \acro{FD}{first differencing}
    \acro{RE}{random effects}
    \acro{MC}{mixed coefficients}
    \acro{IV}{instrumental variable}
    \acro{OLS}{Ordinary Least Squares}
    \acro{POLS}{Pooled Ordinary Least Squares}
    \acro{Wilcoxon test}{Wilcoxon rank distribution test}
    \acro{US}{United States}
    \acro{EU}{the European Union}
    \acro{HSR}{Hart-Scott-Rodino Act of 1976}
    \acro{DOJ}{Department of Justice}
    \acro{SSNIP}{small but significant non-transitory increase in price}
    \acro{IND}{Investigational New Drug}
    \acro{NDA}{New Drug Application}
    \acro{ANDA}{Abbreviated New Drug Application}
    \acro{ODA}{Orphan Drug Act}
    \acro{HAV}{Hepatitis A}
    \acro{HBV}{Hepatitis B}
    \acro{HCV}{Hepatitis C}
    \acro{HIV}{HIV/AIDS}
    \acro{BMS}{Bristol-Myers Squibb}
    \acro{GSK}{Glaxo Smith Kline}
    \acro{ViiV}{ViiV Healthcare}
    \acro{AWP}{Average Wholesale Price}
    \acro{Gilead}{Gilead Sciences}
    \acro{HDHP}{High-Deductible Health Plan}
    \acro{CDHP}{Consumer-Driven Health Plan}
    \acro{POS}{Point of Service}
    \acro{IP}{Intellectual Property}
    \acro{GAFAM}{Google, Amazon, Facebook, Apple, and Microsoft}
    \acro{IO}{Industrial Organization}
    \acro{IC}{Indifference Curve}
    \acro{ISP}{Internet Service Provider}
    \acro{Mbps}{megabit per second}
    \acro{AI/AN}{American-Indian/Alaskan-Native}
    \acro{MMA}{Modern Merger Analysis}
    \acro{CPI}{Consumer Price Index}
    \acro{OTC}{over-the-counter}
    \acro{JJ}{Johnson and Johnson}
    \acro{MS}{multiple sclerosis}
    \acro{CNS}{central nervous system}
    \acro{ADHD}{attention deficit hyper activity disorder}
    \acro{ATC}{Anatomical Therapeutic Chemical}
    \acro{NDC}{National Drug Code}
    \acro{HCP}{health care provider}
    \acro{HCF}{Healthcare Connect Fund}
    \acro{Telecom}{Telecommunications Program}
    \acro{RUCC}{Rural-Urban Continuum Code}
    \acro{FCC}{Federal Communications Commission}
    \acro{BO1}{Best of 1}
    \acro{BO3}{Best of 3}
    \acro{BO5}{Best of 5}
    \acro{CT}{counter-terrorist team}
    \acro{T}{terrorist team}
    \acro{CS:GO}{Counter Strike: Global Offensive}
    \acro{LAN}{Local Area Network}
    \acro{META}{Most Effective Tactic Available}
    \acro{AWP}{Arctic Warfare Police}
    \acro{MPLS}{Multiprotocol Label Switching}
    \acro{ISDN}{Integrated Services Digital Network}
    \acro{USAC}{Universal Service Administrative Company}
    \acro{QALY}{quality-adjusted life years}
    \acro{SPARCS}{Statewide Planning and Research Cooperative System}
    \acro{PCa}{prostate cancer}
    \acro{PSA}{prostate-specific antigen}
    \acro{CDC}{Centers for Disease Control and Prevention}
    \acro{NBS}{Nash Bargaining Solution}
    \acro{FOC}{first order condition}
    \acro{RHC}{Rural Health Care}
    \acro{USF}{Universal Service Fund}
    \acro{NIH}{National Institute of Health}
    \acro{FRC}{Federal Radio Commission}    
    \acro{GMM}{Generalized Method of Moments}
    \acro{CSR}{Corporate Social Responsibility}
    \acro{LBW}{Low Birth Weight}
    \acro{VLBW}{Very Low Birth Weight}
    \acro{ELBW}{Extremely Low Birth Weight}
    \acro{ER}{Emergency Room}
    \acro{SES}{socioeconomic status}
\end{acronym}

\title{Birthweight Declined During the Pandemic and It Is Falling Further Post-pandemic}

\author{Maysam Rabbani\thanks{Department of Economics, Feliciano School of Business, Montclair State University, NJ, USA.}
\and
Elijah Gervais\footnotemark[1]}

\maketitle

\begin{abstract}

Recent literature reports mixed evidence on whether birthweight has decreased during the pandemic. In this paper, we use New York's hospital inpatient discharge data and contribute to this ongoing debate in multiple ways. First, we corroborate that birthweight has declined during the pandemic by 7g (grams). Second, we provide the first empirical evidence that, after the pandemic, not only birthweight has not reverted to the pre-pandemic levels, but it has fallen lower, 17g below the pre-pandemic levels. Third, in the post-pandemic years, mothers who are hospitalized to give birth are 27\% more likely to be at a higher mortality risk and 8\% more likely to have a higher severity of illness. Disruptions to birthweight could have far-reaching consequences for the health, longevity, and well-being of the population. Therefore, understanding the full scope of COVID-19’s influence on birthweight is a vital and timely practice. Future research is needed to test whether our results are driven by true underlying changes in birthweight and complications or by healthcare providers being induced (financially or otherwise) to report birthweight differently.


\end{abstract}

\section{Introduction}
\label{sec:intro}

The COVID-19 pandemic has profoundly shaped the healthcare landscape, leaving damages that are yet to come to light. Among its many impacts, the literature suggests that the pandemic may have affected birthweight, though the evidence is highly mixed \citep{alshaikh2022impact, du2021association, philip2020unprecedented, ranjbar2021changes, tessier2023covid, yao2023covid, yalccin2022effects, briozzo2022impact, mak2023impact, llorca2021results,maki2023no}. A similar mix of evidence is present for premature birth \citep{arnaez2021lack, lin2021covid, briozzo2022impact, ornaghi2022indirect, mak2023impact, ranjbar2021changes, rolnik2021impact, gurol2022obstetric, tessier2023covid, alshaikh2022impact, mor2021impact, son2021coronavirus, badran2021adverse, maki2023no}, gestational age \citep{mor2021impact, gholami2023impact, briozzo2022impact, son2021coronavirus, maki2023no}, hypertensive disorder \citep{son2021coronavirus, mak2023impact}, and whether the C-Section rate has been affected \citep{yalccin2022effects, gurol2022obstetric, mak2023impact, son2021coronavirus, ranjbar2021changes, rabbani_bw2}. The pandemic has potentially affected the rate of induction of labor \citep{gurol2022obstetric, wagner2022perinatal}, obstetric anesthesia \citep{wagner2022perinatal}, gestational diabetes mellitus \citep{ornaghi2022indirect}, and the overall rate of adverse birth outcomes \citep{caniglia2021modest}. During the pandemic, postpartum hospital stays were shorter \citep{wagner2022perinatal}, and more women opted for short stays, who later were more likely to have adverse events \citep{wagner2022perinatal}. During the pandemic, perinatal weight gain increased \citep{nethery2023weight} without a significant change in the rates of newborn mortality \citep{badran2021adverse, ranjbar2021changes}, placental abruption \citep{son2021coronavirus}, intrapartum fetal death rate \citep{mor2021impact}, or postpartum hemorrhage \citep{son2021coronavirus}.

Disruptions to birthweight could have far-reaching consequences for the health, longevity, and well-being of the population \citep{kapral2018associations, yun2021age, sokolovic2013sleep, rush2010birth}. Therefore, understanding the full scope of the pandemic's influence on birthweight is a vital and timely practice. While existing studies have explored the effects of COVID-19 on birthweight and maternal complications during the pandemic, we do not know if the effects reverted to pre-pandemic levels or if there are lingering effects. 

Our primary goal was to answer this question by examining the following outcome measures: birthweight, \ac{LBW} rate, the probability of being discharged home after giving birth, mortality risk, \ac{SOI}, and birth upon \ac{ER} visit. A key advantage of our study is that we separately measured the effect both during and after the pandemic.

By analyzing New York’s hospital inpatient discharge data, we showed that not only has birthweight not rebounded to pre-pandemic levels, but it has continued to decline, raising concerns about the trajectory of maternal and neonatal health. We also reported a sharp increase in the \ac{SOI} and mortality risk.

\section{Methods}
\label{sec:methods}

\subsection{Data}

We used the \ac{SPARCS} data -- New York's hospital inpatient discharge data. It is an administrative database maintained by the New York State Department of Health, capturing detailed, patient-level discharge data from all hospitals and licensed ambulatory surgery and outpatient services in the state. \ac{SPARCS} is at the level of the episode of care, meaning that there is one observation in the data set for each birth case, even if there are several claims associated with it. 

\ac{SPARCS} includes extensive information on demographics, diagnoses, procedures, charges, and discharge status. For birthweight, SPARCS provides near-comprehensive coverage of hospital deliveries in New York, making it a valuable resource for the purpose of our analysis. \ac{SPARCS} data includes maternal and infant characteristics, such as maternal age, race, ethnicity, insurance type, ZIP code, and measures of mortality risk and \ac{SOI}. Infant data includes birthweight, which is key to our analysis. Taking the pandemic as an exogenous shock, \ac{SPARCS} data would enable a quasi-experimental design. A limitation of the data, however, is the absence of individual patient identifiers. So, a panel study is infeasible, and a repeated cross-section approach must be adopted. 

\ac{SPARCS} data was not limited to birth cases. Therefore, the raw sample, spanning 2012-2022, had over 20 million observations. After limiting the sample to birth cases with valid records for all the dependent and independent variables of interest (more in section \ref{sec:results}), a net sample of 2,484,552 birth observations was utilized for analysis. Our exclusion criteria consisted of dropping observations with missing or invalid values for \ac{SOI}, age group, gender, and healthcare facility indicator. Length of stay during the hospitalization was recorded as a numeric in days. The only non-numeric recorded value for length of stay was ``120 +'' which we dropped to keep numeric values. The above steps led to the omission of 0.2\% of the sample, resulting in a net sample of 2,478,149 observations.

\subsection{Outcome and independent variables}

We investigated the potential effect of the pandemic and post-pandemic periods on six outcome measures of interest: (1) newborn birthweight, (2) the probability of \ac{LBW} (weight under 2500g), (3) the probability of being discharged home after delivery, as opposed to other dispositions such as critical access hospital, nursing facility, hospice, deceased, etc., (4) the probability of having a moderate or higher \ac{SOI}, (5) the probability of having a moderate or higher mortality risk, and (6) the probability that the mother was admitted to the \ac{ER} in order to give birth. The combination of these six measures would elucidate if birthweight changed, and if mothers and babies were more likely to face complications during or after the pandemic.

There were two main independent variables of interest: a binary indicator for the pandemic era (year 2020), and a binary indicator for the post-pandemic era (2021-2022). In addition, our set of covariates included newborn gender; race; ethnicity; all possible double and triple interactions of gender, race, and ethnicity; binary indicators for the first, second, and third source of payment; and year in which the case was reported.

\subsection{Hypotheses}

To evaluate the impact of the COVID-19 pandemic on neonatal outcomes, we tested five hypotheses addressing both birthweight and birth-related complications. These were structured to identify deviations from historical trends, capture pandemic-period effects, and assess post-pandemic recovery. Each hypothesis reflected a specific empirical claim that informed whether observed disruptions were temporary or persistent.

\hypothesis{H1: Absent the effect of the pandemic, birthweight is stable over time.}

This hypothesis established a baseline by testing for the presence of any systematic year-to-year variation in birthweight after controlling demographics. Confirming this assumption was necessary to interpret any deviation during the pandemic period as a genuine shock rather than noise or secular drift. If a pre-pandemic trend was detected, we would treat that trend as the counterfactual and assess whether the pandemic and post-pandemic periods represent statistically significant deviations from it. In that case, our estimates would reflect departures from the underlying trajectory rather than from a flat baseline.

\hypothesis{H2: Birthweight declined during the pandemic.}

This hypothesis implied that pandemic-induced disruptions (contracting COVID-19, limited access to care, elevated maternal stress, economic instability, etc.) led to measurable reductions in fetal growth. A statistically significant drop in birthweight during the pandemic would provide direct evidence of the biological and social consequences of the crisis.

\hypothesis{H3: Post-pandemic, birthweight reverted to pre-pandemic levels.}

This tested whether the effects observed during the pandemic, if any, were transitory. A return to pre-pandemic birthweight levels would suggest that the adverse effects did not persist beyond the acute phase of the crisis. Rejection of this hypothesis would point to lasting damage to maternal or fetal health environments.

\hypothesis{H4: The pandemic caused birth complications in the form of changes to the probabilities of being discharged home, mortality risk, \ac{SOI}, and birth upon ER admission.}

This hypothesis examined broader disruptions to neonatal care. Changes in discharge patterns, mortality, \ac{SOI}, or increased reliance on \ac{ER} admissions during delivery would indicate systemic strain on the healthcare infrastructure, with potential consequences for maternal and infant outcomes.

\hypothesis{H5: All the birth complications caused by the pandemic disappeared post-pandemic.}

This final hypothesis asked if the system fully recovered after the pandemic. Evidence of lingering disruptions in discharge status, \ac{SOI}, or mortality would imply persistent structural weaknesses or altered care-seeking behavior.

If all five hypotheses were confirmed, the inference would be that the pandemic introduced sharp but temporary shocks, with no lasting consequences beyond its active period. In contrast, a rejection of \textit{H3} or \textit{H5} would suggest that the pandemic had persistent effects on either biological outcomes or healthcare delivery systems. These findings would carry direct implications for the design of post-crisis health policy and perinatal care protocols\citep{lavin2019zika, abrahams2024disaster, worldbank2020drf, undp2023health, hhs2021toolkit}.

\subsection{Statistical model}

We implemented two specifications to test the above hypotheses. In specification 1, we introduced a binary (dummy) control for the pandemic (equal to $1$ in 2020 and $0$ otherwise) and another binary control for the post-pandemic (equal to $1$ in years 2021-2022 and $0$ otherwise). The former captures the difference in the outcome measures between the pandemic era and pre-pandemic, and the latter captures the difference in the outcome measures between the post-pandemic and pre-pandemic. The inclusion of these variables alongside an annual time trend allowed us to isolate the secular effect of the pandemic and post-pandemic as shocks to the outcome measures. In specification 2, we introduced year fixed-effects, where 2012 served as the omitted group, and 10 binary variables captured annual changes during 2013-2022. 

The two specifications complemented one another in the following manner: specification 1 provided a straightforward statistical test for whether the pandemic and post-pandemic eras shifted the outcome measures, whereas specification 2 provided a refined picture of the evolution of the outcome measures as the population went into and out of the pandemic. Specification 1 was used in the baseline regression estimation (Table \ref{tab:reg1}) and specification 2 was the basis for the event study (Figure \ref{fig:event}).

We used the following model for specification 1:

\begin{equation}
    LHS_{it} = \beta_0 + \beta_1.Year_{it} + \beta_2.Pandemic_{it} + \beta_3.PostPandemic_{it} +  \boldsymbol{XB} + \epsilon_{it}
    \label{eq:1}
\end{equation}

Borrowing from the literature \citep{rabbani_bb1}, we used the following model for specification 2:

\begin{equation}
    LHS_{it} = \alpha_0 + \sum_{y=2013}^{2022}[\alpha_y.I^y_{it}] + \boldsymbol{XA} + e_{it}
    \label{eq:2}
\end{equation}

In the above models, $LHS_{it}$ represents the outcome measure. For birthweight, a continuous outcome measure, we implemented a \ac{POLS} approach. We used a logit model for all other outcome measures that are binary. $Year_{it}$ is a continuous measure for the hospitalization year with index $t$ for year and index $i$ for the individual. $Pandemic_{it}$ is equal to $1$ in 2020 onward and $0$ otherwise. $PostPandemic_{it}$ is equal to $1$ in 2021-2022 and $0$ otherwise. $\boldsymbol{XB}$ and $\boldsymbol{XA}$ are the vectors of covariates and their coefficients, including newborn's gender, race, and ethnicity; all double and triple interactions of gender, race, and ethnicity; binary insurance plan indicators for the first, second, and third source of payment; a binary indicator equal to $1$ if the \ac{SOI} is anything above ``minor'' and $0$ if it is ``minor''; and a binary indicator equal to $1$ if the mortality risk is anything above ``minor'' and $0$ if it is ``minor''. $I^y_{it}$ is a binary indicator equal to $1$ in year $y\in [2013,2022]$ and $0$ otherwise.  $\epsilon_{it}$ is the error term with a standard normal distribution for birthweight (\ac{POLS}) and logistics distribution for all other outcome measures (logit). We clustered the standard errors at the healthcare facility level to account for the heteroskedasticity that could arise because of facility-level variations in health outcomes.

\section{Results}
\label{sec:results}

Table \ref{tab:summary} reports the summary statistics. 8\% of the births occurred during, and 17\% after the pandemic. 8\% of the sample were identified as \ac{LBW}, 2\% of the births were upon \ac{ER} admission, and 95\% of the patients were discharged home. The sample consisted of 47\% white, 14\% black, and 39\% other races. 15\% of the sample were Hispanic, 72\% are non-Hispanic, and 13\% did not identify. The mortality risk was minor in 97\%, moderate in 2\%, major in 1\%, and extreme in less than 1\% of the cases. Likewise, the \ac{SOI} was minor in 74\%, moderate in 18\%, major in 7\%, and extreme in 1\% of the cases. The majority of the sample used Medicaid (50\%), private health insurance (24\%), or Blue Cross Blue Shield (19\%). A net sample of 2,478,149 observations provided sufficient statistical power to reliably study the outcome measures.

\insertitem{TABLE 1}

\begin{figure}[ht]
    \centering
    \includegraphics[width=1\linewidth]{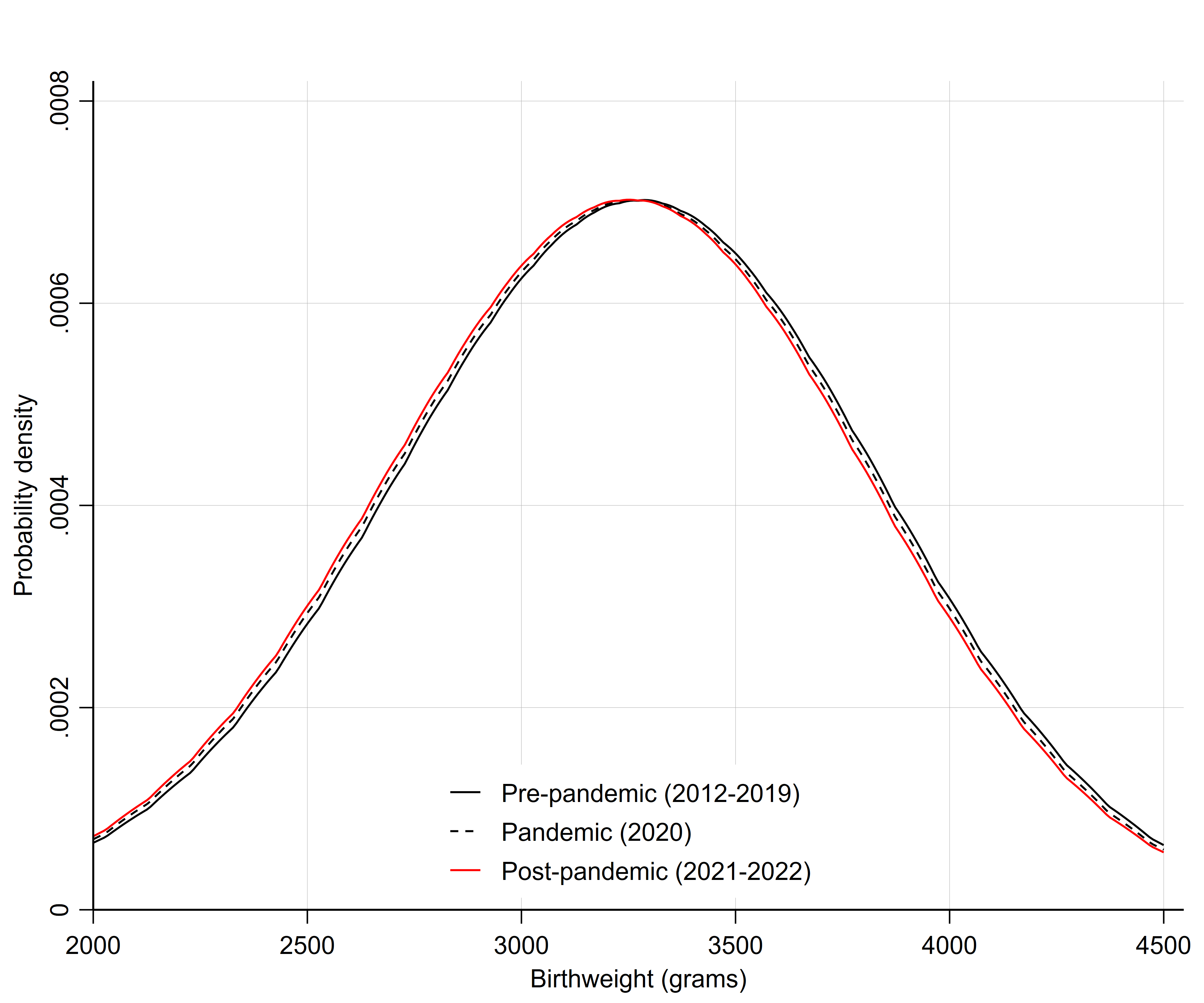}
    \caption{Kernel density of birthweight pre-pandemic (2012–2019), during the pandemic (2020), and in the post-pandemic years (2021–2022). Across the three time intervals, a small but consistent shift to the left is visible.}
    \label{fig:bweight_density}
\end{figure}

Figure \ref{fig:bweight_density} shows a kernel density distribution fit for birthweight. Three curves are overlaid. The solid black curve shows the pre-pandemic, the dashed black curve displays the pandemic, and the red curve shows the post-pandemic kernel density distribution of birthweight. If there were truly no shifts in birthweight, the three density distributions should be indistinguishable. However, the trends are mis-aligned: starting with the pre-pandemic (solid black), the pandemic trend (dashed black) is shifted to the left, and the post-pandemic (red) is further to the left. In other words, the kernel density distribution suggests a small but visible and universal shift of the distribution to the left, providing suggestive evidence that birthweight has been on the decline during and after the pandemic.

\insertitem{FIGURE 1}

\begin{figure}[ht]
    \centering
    \includegraphics[width=\linewidth]{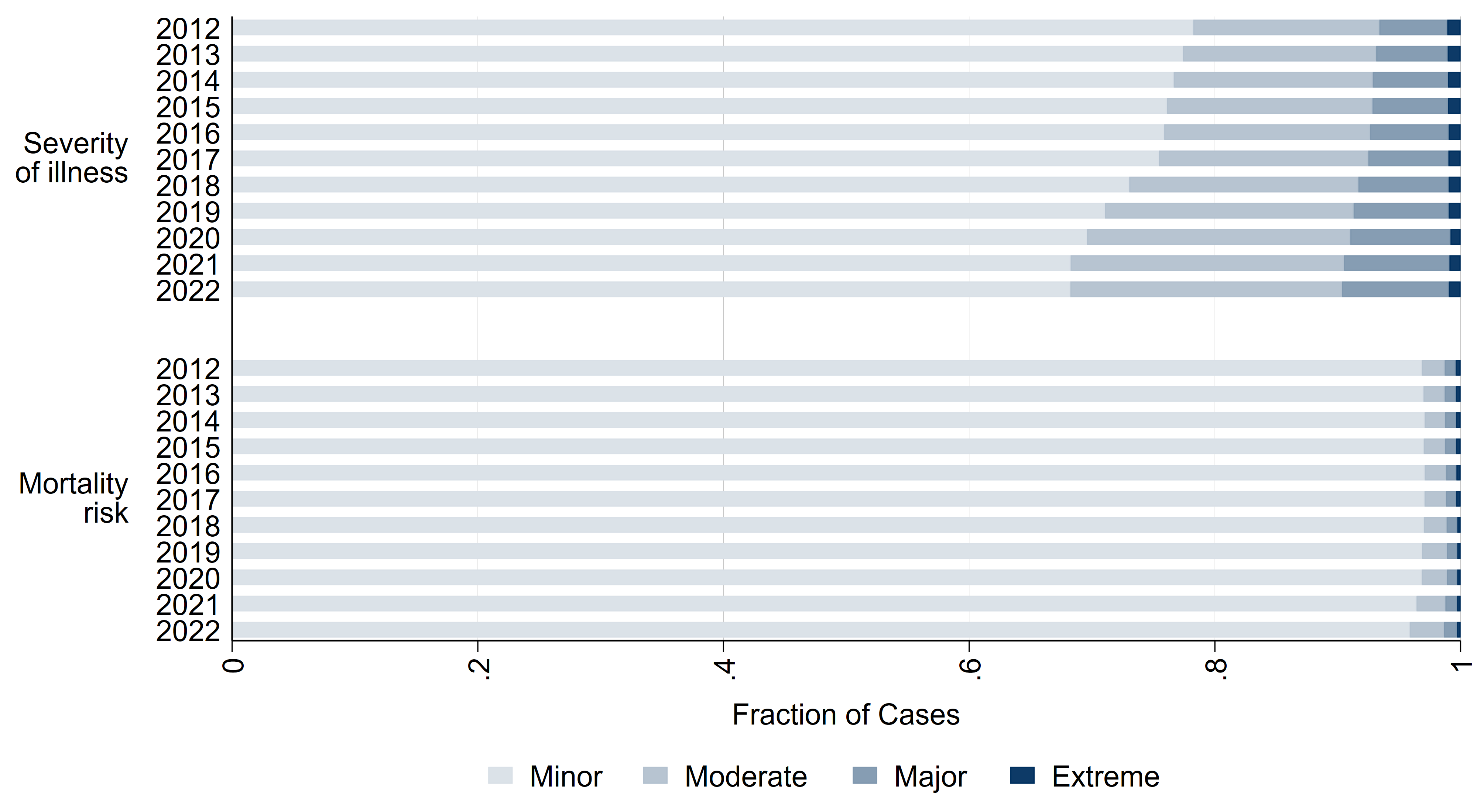}
    \caption{Severity of illness (left) and mortality risk (right) composition during 2012-2022. Each bar shows the fraction of cases that were at minor, moderate, major, or extreme levels of intensity. Intensity levels are color-coded from the lightest for minor to darkest for extreme.}
    \label{fig:soi_mortality}
\end{figure}

Figure \ref{fig:soi_mortality} shows the trends of \ac{SOI} (top) and mortality risk (bottom). \ac{SPARCS} data reports \ac{SOI} and mortality risk in a similar four-tier classification: minor, moderate, major, and extreme. Figure \ref{fig:soi_mortality} shows the fraction of birth cases in each year that were reported to have a minor (lightest color), moderate, major, or extreme (darkest color) level of \ac{SOI} or mortality risk. For both \ac{SOI} and mortality risk, most birth cases were marked with minor intensity. However, the fraction with minor intensity is declining, and it appears to be replaced with moderate and major levels of intensity. The fraction of observations with extreme intensity level appears to be stable over time for both \ac{SOI} and mortality risk.

\insertitem{TABLE 2}

\begin{figure}[ht]
    \centering
    \includegraphics[width=.7\linewidth]{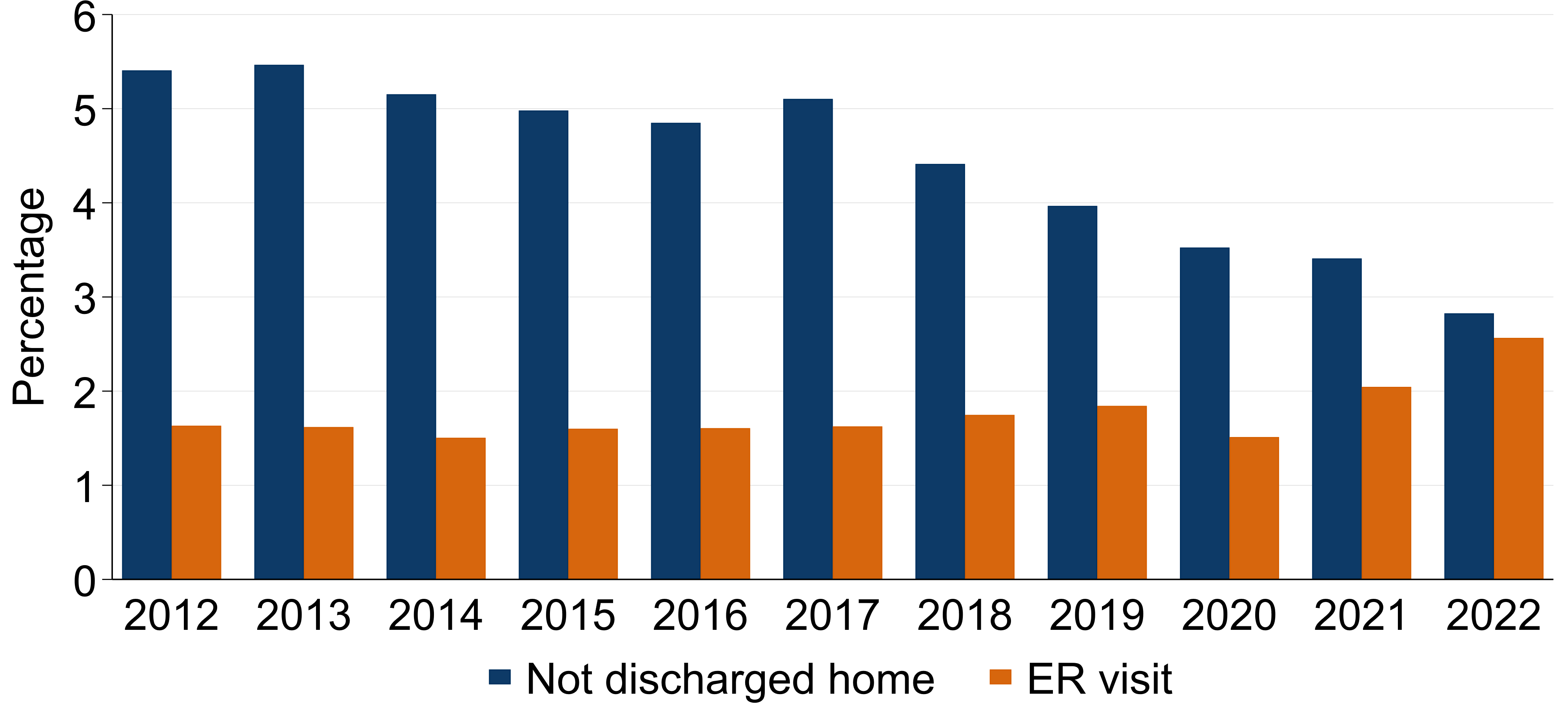}
    \caption{Percentage of birth cases in which the patient wasn't discharged to their home (light gray) and percentage of birth cases in which the patient has been admitted to the \ac{ER} (dark gray).}
    \label{fig:discharge_er}
\end{figure}

Table \ref{tab:disposition} reports patient disposition data by year. Disposition contains information about patient's health status and destination at the time of being discharged. Among the many dispositions, arguably the best possible outcome is being discharged home under self-care, which makes up the overwhelming majority of the sample. There is no clear ordering of the remaining patient disposition categories. For instance, while it is clear that being ``expired'' is worse than leaving ``against medical advice'', it is unclear whether being discharged to a short-term hospital is better or worse than an inpatient rehabilitation facility. Hence, we proceed with the assumption that anything other than being discharged home is an inferior outcome than being discharged home, and we use the fraction of patients who were not discharged home as a measure of health outcomes. This is best shown in Figure \ref{fig:discharge_er}.

\insertitem{FIGURE 2}

Figure \ref{fig:discharge_er} reports two statistics: the fraction of birth cases that were admitted to the \ac{ER}, and the fraction of patients that were not discharged to their homes. A reduction in both measures could indicate improvement because the best case for admission is when the patient has scheduled it (as opposed to an unscheduled \ac{ER} visit) and the best discharge case is being sent home (as opposed to being discharged to another healthcare facility). Two observations stand out. There is a steady decline in the light gray bar, indicating that a growing number of patients are being discharged home. Meanwhile, a growing number of patients are admitted to the \ac{ER}. During and after the pandemic, the fraction of \ac{ER} admissions had a sudden dip in 2020 followed by a rapid rise in 2021-2022.

\insertitem{FIGURE 3}

\begin{figure}[ht]
    \centering
    \includegraphics[width=1\linewidth]{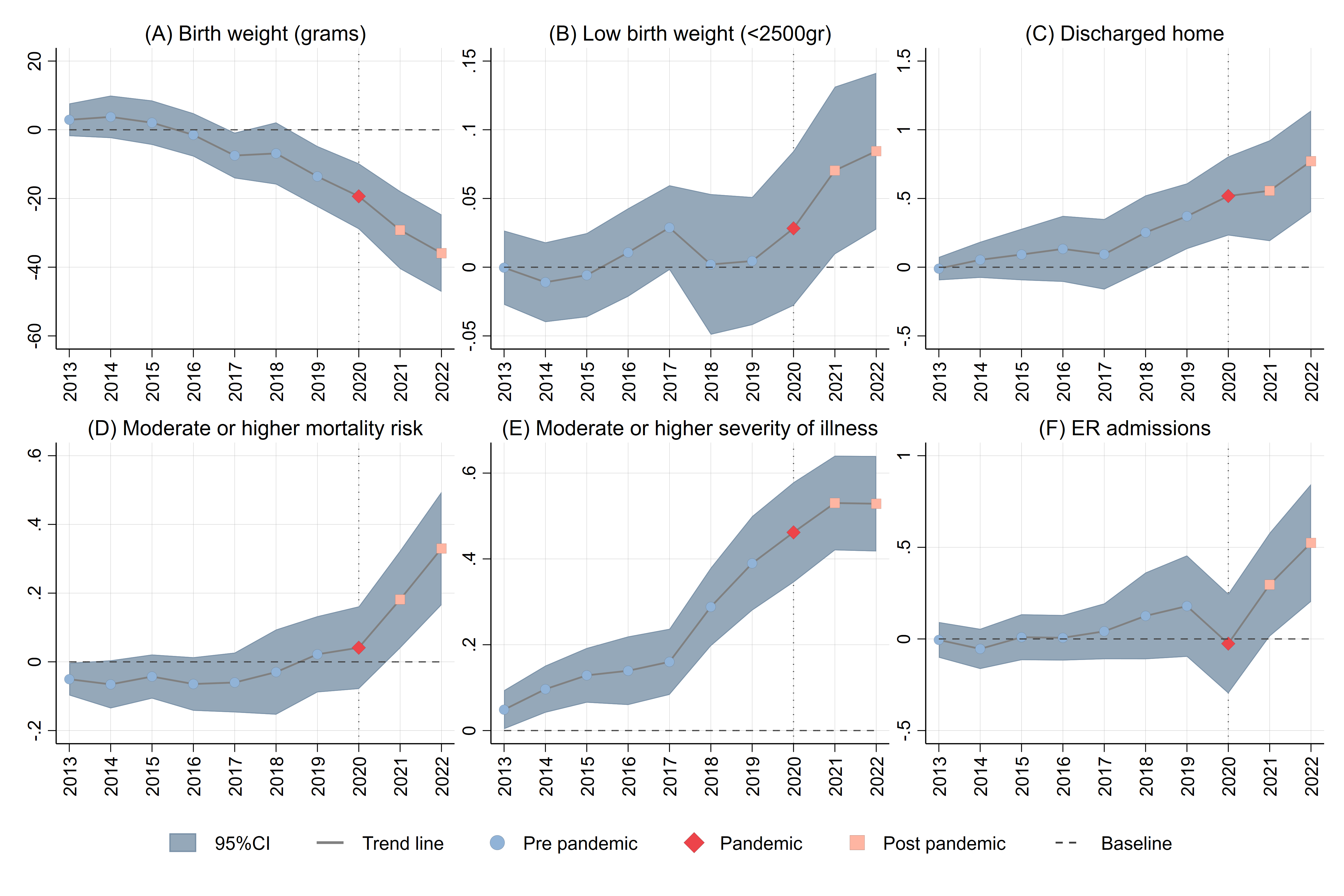}
    \caption{event study trends for birthweight, \ac{LBW}, being discharged home, moderate or higher risk of mortality, moderate or higher \ac{SOI}, and birth upon \ac{ER} admission. This figure is created using specification 2.}
    \label{fig:event}
\end{figure}

Table \ref{tab:reg1} reports the baseline regression estimates using specification 1. The table reports the estimated coefficients, $\beta$. Readers may recover the odds ratio using $OR=\exp({\beta})$. Standard errors, clustered at the healthcare facility level, are reported in parenthesis. Column 1 implements \ac{POLS}, and columns 2-6 use logit. Figure \ref{fig:event} illustrates the event study results using the coefficients of years in specification 2. The pandemic period is highlighted in red, post-pandemic in yellow, and pre-pandemic in blue. A 95\% confidence interval envelops the trend lines to facilitate the interpretation of statistical significance.

\insertitem{TABLE 2}

Column 1 shows that birthweight is not stable over time. Instead, it is declining by 2.17g per year. During the pandemic, birthweight dropped 6.98g below the pre-pandemic levels. Post-pandemic, birthweight did not revert to the pre-pandemic levels. Instead, it fell by an additional 9.95g beyond the pandemic era, resulting in post-pandemic birthweight to be 16.93g below the pre-pandemic levels. Not only does this reject $H3$, which is concerning, but it raises the concern that the decline in birthweight is accelerating post-pandemic. Panel A in Figure \ref{fig:event} supports this result. While there is partial evidence of a decline that began in 2019, we observe a clear drop in birthweight during the pandemic, followed by a greater downward trend post-pandemic.

\insertitem{FIGURE 4}

Column 2 confirms that \ac{LBW} is stable over time and there is no significant change during the pandemic. However, babies born post-pandemic are 6.00\% more likely to have \ac{LBW}. Panel B in Figure \ref{fig:event} supports this, showing that \ac{LBW} was stable throughout the years, only to begin rising in 2021 and further in 2022.

Column 3 shows that each year, patients are 4.77\% more likely to be discharged home than the year prior. This finding is encouraging and hints at a steady improvement in health outcomes because being discharged home is the best possible patient disposition. During the pandemic, patients were 18.1\% more likely to be discharged home. A higher frequency of discharge to home at the peak of the pandemic could be a natural response of the healthcare system: hospitals were hot spots of COVID-19 contagion, and erring on the side of discharging early and to home would be a reasonable adjustment. However, being discharged home further intensified post-pandemic: patients were 25.1\% more likely to be discharged home post-pandemic than pre-pandemic, which is 7.00\% higher than the pandemic years. Panel C in Figure \ref{fig:event} raises the possibility that discharge home has been steadily rising since 2018. So, we cannot necessarily attribute it to the pandemic, and we leave it to a future study to delve deeper into the underlying causes.

As shown in Column 4, mortality risk is stable over the pre-pandemic years, and there is no statistically significant change during the pandemic. However, after the pandemic, patients are 26.7\% more likely to have a higher mortality risk. Panel D in Figure \ref{fig:event} strongly supports this: none of the annual coefficients are significant during 2013-2020, but there is a sharp spike in the mortality risk in 2021 that escalated in 2022. Uncovering the underlying reasons is a critical question that is beyond the purview of this study.

As shown in Column 5, \ac{SOI} has been rising annually by 4.87\% pre-pandemic. Besides this annual increase, the pandemic led to an additional 8.58\% increase that has remained stable post-pandemic (adjusting from 8.58\% to 8.03\%). However, Panel E in Figure \ref{fig:event} suggests that it may not be attributable to the pandemic. The trend seems to have begun in 2018. We leave it to a future study to validate this possibility and explore the potential events in 2018 that could explain this change of pace.

Column 6 shows that a stable fraction of the pre-pandemic birth cases were upon \ac{ER} admission. But it declined by 19.5\% during the pandemic. Given the tendency during the pandemic to stay away from urgent healthcare facilities, a lower \ac{ER} admission would be reasonable. What is odd, however, is that after the pandemic, \ac{ER} admissions to give birth rose 20.4\% above the pre-pandemic rates. In other words, post-pandemic people are less likely than pre-pandemic to have their births planned and scheduled and, hence, more likely to be admitted to the \ac{ER} or other non-elective settings. This remains to be further validated in future studies.

We observe discrepancies in birthweight and complications based on gender, race, and ethnicity. Notably, female newborns are 128.7g lighter than male, black newborns are 203.6g lighter than white, and Hispanic newborns are 70.33g lighter than non-Hispanic newborns. These statistics are not concerning per se because different genders, races, and ethnicities could be born at different normal weights. However, a finding that is potentially concerning and noteworthy is that black newborns are 39.3\% more likely to face a higher mortality risk and 29.0\% more likely to have a higher \ac{SOI}. While this is beyond the purview of this study, we encourage future work to investigate this. More importantly, it is a vital next step to test if the pandemic has had distinct effects on different demographics.

\section{Discussions}
\label{sec:discussions}

Using the estimates in the previous section, we can report the hypothesis test results as follows. The results reject $H1$: birthweight is not stable over time. Even before the pandemic, birthweight was falling by 2.17g annually. The results confirm $H2$: birthweight declined during the pandemic by 6.98g. The results reject $H3$: birthweight did not revert to pre-pandemic levels. Instead, it fell by an additional 9.95g. The results confirm $H4$: the pandemic led to an 8.58\% increase in the \ac{SOI}, and reduced births upon \ac{ER} admissions by 19.5\%. The results reject $H5$: not only did birth complications not revert to pre-pandemic levels, but we also documented the appearance of post-pandemic trends that were not present during the pandemic. Specifically, there is a sharp post-pandemic rise in the mortality risk and \ac{SOI} that was not there during the pandemic.

While we documented the resulting birth outcomes during and after the pandemic, we leave it to future studies to further explore the underlying reasons. We conjecture that one of the following possibilities is driving the results. First, COVID-19 and the health complications that it has caused could have resulted in a permanent change in birthweight. In our view, this is a possible but unlikely scenario. Second, it is well understood that certain health behaviors such as annual checkups, preventive care, and regular contact with primary care physicians were disrupted during the pandemic. If such a change of habits has been lingering post-pandemic, pregnant women may have fallen behind in the utilization of maternal screenings and reduced their contact with their obstetrician/gynecologist, and this could be driving poorer health outcomes post-pandemic \citep{habibov2017effect, almahmoud2023three}. A third possibility is that birth outcomes have not changed, and what has changed is the way the healthcare system is reporting and documenting patient data. Financial motives are known to influence the course of medical decisions and reporting \citep{hansen2005changes}. For example, a study found that upon the introduction of a capitated payment in Germany, over 12000 birth cases were upcoded to artificially report a lower birthweight to increase the reimbursement \citep{jurges2013first}. While we encourage future research to look further into all the above possibilities, we avoid drawing any conclusions regarding the underlying causes.

\section{Limitations}
\label{sec:limitations}

Our study is limited in the following ways. First, we haven't ruled out the possibility that latent factors are affecting birthweight and birth outcomes. The literature has shown that other factors are in play such as regulatory and reimbursement reforms \citep{jeungimproving, webster2022state} and maternal socioeconomic and behavioral characteristics \citep{perkowski2024co21, bolbocean2022ee596, zhu2016prevalence, smolen2015development, rainham2007differences}. 

Second, our analysis remains a correlation study. That is, we found statistically significant changes in birthweight and complications during and after the pandemic. However, we cannot claim causality, and it is yet to be investigated whether the changes were caused by the pandemic or by latent concurrent events. Furthermore, we do not know if the true weight of newborns has changed or if it is being upcoded.

Third, our cutoff points do not perfectly match the exact start and end date of the pandemic. This limitation is imposed by the data because the only time measure that is reported is the year. Fourth, our study is limited to the state of New York. It is imperative to test the same question in other geographies to understand if it is an isolated pattern or a universal phenomenon.

\section{Conclusions}
\label{sec:conclusions}

As society begins to view the pandemic in retrospect, there is a tendency to assume that its health impacts are behind us and that disruptions to birthweight and maternal health have naturally resolved. However, our findings challenge this assumption. To our knowledge, this is the first paper that investigates birth outcomes separately for the pandemic years and the post-pandemic era. We documented several important results. First, birthweight has fallen during the pandemic, and not only has it not recovered after the pandemic, but it is also on a downward spiral. Second, we found concerning results regarding birth complications: in the years following the pandemic, newborns face a higher degree of mortality risk and severity of illness, at levels that were unprecedented even during the pandemic. Third, we find that birthweight has been on a downward path for the past decade: each year, newborns are 2.17g lighter than newborns in the year prior. 

In short, the results are highly concerning on three fronts: birthweight has been steadily falling for the past decade, the decline in birthweight that began during the pandemic is intensifying post-pandemic, and the severity of illness and mortality risk have sharply increased post-pandemic. We recognize the limitations of our study and emphasize the need for further research to validate and expand upon these findings. Developed countries tend to face similar challenges with regard to birth and pregnancy complications \citep{papanicolas2024maternal}. Given the importance of this issue and the potential consequences for public health, it is imperative to start a broader conversation. 



\subsection*{Conflicts of interest}

The authors declare that no funds, grants, or other support were received during the preparation of this manuscript. The authors have no competing or conflicting interests to report. 






\pagebreak

\begingroup
\setstretch{.6} 
\scriptsize 
\bibliography{References.bib}
\endgroup

\pagebreak




\section*{Supporting tables and figures}
\label{sec:app_tables_figs}

\begin{table}[ht]
\centering
\caption{summary statistics.}
\label{tab:summary}
\resizebox{.8\columnwidth}{!}{%
\begin{tabular}{@{}lcccccc@{}}
\toprule
Variable & Obs & Mean & SD & Min & Median & Max \\ \midrule
Pandemic   (2020) & 2,478,149 & 0.08 & 0.28 & 0.00 & 0.00 & 1.00 \\
Post-pandemic   (2021 and 2022) & 2,478,149 & 0.17 & 0.37 & 0.00 & 0.00 & 1.00 \\
Year & 2,478,149 & 2016.81 & 3.15 & 2012 & 2017 & 2022 \\
Birthweight (grams) & 2,478,149 & 3209.59 & 585.79 & 100 & 3200 & 12000 \\
Birth upon \ac{ER} visit & 2,478,149 & 0.02 & 0.13 & 0.00 & 0.00 & 1.00 \\
Low birthweight & 2,478,149 & 0.08 & 0.27 & 0.00 & 0.00 & 1.00 \\
Discharge home & 2,478,149 & 0.95 & 0.21 & 0.00 & 1.00 & 1.00 \\
Female newborn & 2,478,149 & 0.49 & 0.50 & 0.00 & 0.00 & 1.00 \\
\multicolumn{7}{l}{\textbf{Race}} \\
\textit{White} & 2,478,149 & 0.47 & 0.50 & 0.00 & 0.00 & 1.00 \\
\textit{Black} & 2,478,149 & 0.14 & 0.34 & 0.00 & 0.00 & 1.00 \\
\textit{Other} & 2,478,149 & 0.39 & 0.49 & 0.00 & 0.00 & 1.00 \\
\multicolumn{7}{l}{\textbf{Ethnicity}} \\
\textit{Hispanic} & 2,478,149 & 0.15 & 0.35 & 0.00 & 0.00 & 1.00 \\
\textit{Non-Hispanic} & 2,478,149 & 0.72 & 0.45 & 0.00 & 1.00 & 1.00 \\
\textit{Other} & 2,478,149 & 0.13 & 0.35 & 0.00 & 0.00 & 1.00 \\
\multicolumn{7}{l}{\textbf{Risk of mortality}} \\
\textit{Minor} & 2,478,149 & 0.97 & 0.18 & 0.00 & 1.00 & 1.00 \\
\textit{Moderate} & 2,478,149 & 0.02 & 0.14 & 0.00 & 0.00 & 1.00 \\
\textit{Major} & 2,478,149 & 0.01 & 0.09 & 0.00 & 0.00 & 1.00 \\
\textit{Extreme} & 2,478,149 & 0.00 & 0.06 & 0.00 & 0.00 & 1.00 \\
\multicolumn{7}{l}{\textbf{\ac{SOI}}} \\
\textit{Minor} & 2,478,149 & 0.74 & 0.44 & 0.00 & 1.00 & 1.00 \\
\textit{Moderate} & 2,478,149 & 0.18 & 0.39 & 0.00 & 0.00 & 1.00 \\
\textit{Major} & 2,478,149 & 0.07 & 0.25 & 0.00 & 0.00 & 1.00 \\
\textit{Extreme} & 2,478,149 & 0.01 & 0.10 & 0.00 & 0.00 & 1.00 \\
\multicolumn{7}{l}{\textbf{Primary source of payment}} \\
\textit{Blue Cross Blue Shield} & 2,478,149 & 0.19 & 0.39 & 0.00 & 0.00 & 1.00 \\
\textit{Department Of Corrections} & 2,478,149 & 0.00 & 0.01 & 0.00 & 0.00 & 1.00 \\
\textit{Federal, State, Local, VA} & 2,478,149 & 0.01 & 0.10 & 0.00 & 0.00 & 1.00 \\
\textit{Managed Care} & 2,478,149 & 0.02 & 0.15 & 0.00 & 0.00 & 1.00 \\
\textit{Medicaid} & 2,478,149 & 0.50 & 0.50 & 0.00 & 0.00 & 1.00 \\
\textit{Medicare} & 2,478,149 & 0.00 & 0.04 & 0.00 & 0.00 & 1.00 \\
\textit{Miscellaneous} & 2,478,149 & 0.00 & 0.04 & 0.00 & 0.00 & 1.00 \\
\textit{Private health insurance} & 2,478,149 & 0.24 & 0.43 & 0.00 & 0.00 & 1.00 \\
\textit{Self pay} & 2,478,149 & 0.03 & 0.18 & 0.00 & 0.00 & 1.00 \\
\textit{Unknown} & 2,478,149 & 0.00 & 0.01 & 0.00 & 0.00 & 1.00 \\ \bottomrule
\end{tabular}%
}
\end{table}

\begin{table}[ht]
\centering
\caption{patient disposition by year.}
\label{tab:disposition}
\resizebox{\columnwidth}{!}{%
\begin{tabular}{lccccccccccc}
\hline
Patient has been   discharged to & 2012 & 2013 & 2014 & 2015 & 2016 & 2017 & 2018 & 2019 & 2020 & 2021 & 2022 \\ \hline
Home or self care & 232,329 & 227,582 & 227,656 & 222,909 & 220,978 & 215,144 & 210,560 & 209,579 & 199,355 & 198,342 & 201,856 \\
Home with home health services & 9,728 & 9,685 & 9,158 & 8,486 & 8,204 & 8,284 & 7,112 & 5,911 & 4,628 & 4,274 & 3,185 \\
Short-term hospital & 2,314 & 2,091 & 1,968 & 2,012 & 1,932 & 2,085 & 1,729 & 1,837 & 1,714 & 1,707 & 1,713 \\
Expired (dead) & 807 & 807 & 711 & 696 & 670 & 668 & 442 & 471 & 396 & 420 & 413 \\
Cancer center or children's hospital & 172 & 196 & 192 & 171 & 187 & 144 & 124 & 152 & 200 & 265 & 275 \\
Another type not listed & 58 & 98 & 124 & 104 & 66 & 95 & 98 & 45 & 117 & 139 & 73 \\
Left against medical advice & 69 & 125 & 44 & 53 & 40 & 54 & 40 & 63 & 47 & 38 & 37 \\
Court/law enforcement & 21 & 23 & 34 & 26 & 37 & 49 & 44 & 44 & 34 & 30 & 27 \\
Skilled nursing home & 35 & 51 & 44 & 33 & 33 & 38 & 25 & 17 & 21 & 15 & 22 \\
Facility with custodial/supportive care & 29 & 27 & 23 & 18 & 16 & 27 & 24 & 37 & 39 & 36 & 26 \\
Inpatient rehabilitation facility & 13 & 14 & 16 & 26 & 25 & 51 & 31 & 22 & 30 & 18 & 28 \\
Hospice - home & 16 & 18 & 20 & 24 & 25 & 26 & 23 & 29 & 17 & 23 & 24 \\
Critical access hospital & 6 & 9 & 14 & 19 & 14 & 14 & 15 & 16 & 21 & 24 & 25 \\
Medicare certified long term care   hospital & 5 & 2 & 6 & 10 & 5 & 15 & 4 & 7 & 6 & 4 & 3 \\
Hospice - medical facility & 4 & 11 & 6 & 4 & 4 & 6 & 4 & 1 & 8 & 3 & 4 \\
Federal health care facility & 2 & 1 & 2 & 2 & 2 & 5 & 6 & 5 & 0 & 1 & 13 \\
Psychiatric hospital or unit of hospital & 0 & 1 & 6 & 1 & 2 & 7 & 0 & 0 & 2 & 1 & 2 \\
Medicaid certified nursing facility & 0 & 0 & 0 & 0 & 0 & 3 & 0 & 1 & 4 & 2 & 0 \\
Hospota-based Medicare-approved swing bed & 0 & 0 & 1 & 0 & 1 & 0 & 0 & 0 & 0 & 0 & 0 \\ \hline
\end{tabular}%
}
\end{table}

\begin{table}[ht]
\centering
\caption{regression estimates.}
\label{tab:reg1}
\resizebox{\columnwidth}{!}{%
\begin{tabular}{@{}lcccccc@{}}
\toprule
\begin{tabular}[c]{@{}l@{}}Dependent \\ variable\end{tabular} & \begin{tabular}[c]{@{}c@{}}Birth \\ weight\end{tabular} & \begin{tabular}[c]{@{}c@{}}Low \\ birth \\ weight\end{tabular} & \begin{tabular}[c]{@{}c@{}}Discharged\\ home\end{tabular} & \begin{tabular}[c]{@{}c@{}}Moderate\\ or higher \\ mortality risk\end{tabular} & \begin{tabular}[c]{@{}c@{}}Moderate or\\ higher severity \\ of illness\end{tabular} & \begin{tabular}[c]{@{}c@{}}Birth \\ upon \ac{ER} \\ admission\end{tabular} \\
Specification & POLS & Logit & Logit & Logit & Logit & Logit \\
Model & (1) & (2) & (3) & (4) & (5) & (6) \\ \midrule
Year & -2.170*** & 0.00230 & 0.0477** & 0.00449 & 0.0487*** & 0.0290 \\
 & (0.629) & (0.00341) & (0.0198) & (0.00777) & (0.00839) & (0.0194) \\
During pandemic (2020) & -6.981** & 0.0143 & 0.181** & 0.0567 & 0.0858*** & -0.195*** \\
 & (3.082) & (0.0162) & (0.0885) & (0.0402) & (0.0313) & (0.0583) \\
Post-pandemic (2021, 2022) & -16.93*** & 0.0600*** & 0.251** & 0.267*** & 0.0803** & 0.204* \\
 & (3.956) & (0.0167) & (0.118) & (0.0722) & (0.0396) & (0.109) \\
Female newborn & -128.7*** & 0.221*** & 0.107*** & -0.280*** & -0.155*** & -0.211*** \\
 & (1.257) & (0.0129) & (0.0189) & (0.0161) & (0.00703) & (0.0218) \\
\multicolumn{7}{l}{\textbf{Race indicators (baseline: white)}} \\
\textit{Black} & -203.6*** & 0.649*** & -0.00189 & 0.393*** & 0.290*** & -0.0330 \\
\textit{} & (8.276) & (0.0613) & (0.142) & (0.117) & (0.0526) & (0.132) \\
\textit{Other} & -131.1*** & 0.300*** & 0.202 & 0.0435 & 0.0995* & 0.0499 \\
\textit{} & (10.29) & (0.0663) & (0.162) & (0.135) & (0.0537) & (0.151) \\
\multicolumn{7}{l}{\textbf{Ethnicity indicators (baseline:   non-Hispanic)}} \\
\textit{Hispanic} & -70.33*** & 0.132 & 0.281 & -0.216 & -0.0477 & 0.296 \\
\textit{} & (11.31) & (0.0852) & (0.209) & (0.156) & (0.0869) & (0.192) \\
\textit{other} & 11.47 & -0.0853 & 0.0241 & -0.324** & -0.0411 & -0.984*** \\
 & (11.08) & (0.0813) & (0.261) & (0.154) & (0.0789) & (0.220) \\
\multicolumn{7}{l}{\textbf{First source of payment (baseline:   Medicaid)}} \\
\textit{Blue Cross Blue Shield} & 57.95*** & -0.178*** & 0.686*** & -0.242*** & -0.117** & -0.548*** \\
\textit{} & (9.371) & (0.0468) & (0.134) & (0.0716) & (0.0457) & (0.115) \\
\textit{Department of corrections} & -92.23** & -0.368 & -2.089*** & -0.189 & 0.108 & -0.306 \\
\textit{} & (35.90) & (0.375) & (0.371) & (0.393) & (0.150) & (0.318) \\
\textit{Federal, State, Local, VA} & 71.57*** & -0.197 & 0.231 & -0.336 & -0.213** & -0.465** \\
\textit{} & (19.64) & (0.134) & (0.202) & (0.342) & (0.0993) & (0.199) \\
\textit{Managed care} & 24.67 & 0.0316 & 0.697*** & -0.117 & -0.0300 & 0.100 \\
\textit{} & (26.62) & (0.179) & (0.199) & (0.298) & (0.125) & (0.365) \\
\textit{Medicare} & 21.79 & -0.0902 & 0.484** & -0.333 & -0.0615 & -0.743*** \\
\textit{} & (26.05) & (0.141) & (0.240) & (0.218) & (0.189) & (0.261) \\
\textit{Miscellaneous} & -14.63 & 0.250** & -0.0516 & 0.166 & 0.00878 & -0.521* \\
\textit{} & (33.32) & (0.116) & (0.291) & (0.275) & (0.128) & (0.293) \\
\textit{Private health insurance} & 36.04*** & -0.172*** & 0.944*** & -0.353*** & -0.130** & -0.562*** \\
\textit{} & (10.40) & (0.0573) & (0.191) & (0.112) & (0.0550) & (0.168) \\
\textit{Self pay} & 38.76*** & -0.206*** & 0.0417 & -0.291*** & -0.109** & -0.417*** \\
\textit{} & (11.10) & (0.0528) & (0.111) & (0.0911) & (0.0549) & (0.136) \\
\textit{Unknown} & -58.51 & 0.150 & 1.564* & -0.110 & 0.333** & -0.591 \\
 & (39.48) & (0.294) & (0.906) & (0.297) & (0.134) & (0.384) \\
Constant & 7,645*** & -7.132 & -93.53** & -12.24 & -99.41*** & -62.73 \\
 & (1,271) & (6.874) & (39.92) & (15.66) & (16.89) & (39.13) \\
Observations & 2,478,149 & 2,478,149 & 2,478,104 & 2,478,139 & 2,478,149 & 2,477,207 \\ \bottomrule
\end{tabular}%
} \\
{\raggedright
\fontsize{6}{6.5}\selectfont 
Standard errors clustered at the healthcare facility level are reported in parenthesis. Although not reported for brevity, all models have binary controls for all possible interactions of gender, race, and ethnicity, binary controls for the second source of payment, and binary controls for the third source of payment. *** $p<0.01$, ** $p<0.05$, * $p<0.1$. \par}
\end{table}

\end{document}